\documentclass[manuscript]{emulateapj}
\usepackage{amsmath}

\shorttitle{Constraining the CMB Optical Depth Through the Dispersion Measure of Cosmological Radio Transients}
\shortauthors{Fialkov \& Loeb}

\begin{document}

\title{Constraining the CMB Optical Depth Through the Dispersion Measure of Cosmological Radio Transients}

\author{Anastasia Fialkov \& Abraham Loeb}
\affil{Institute for Theory and Computation, Harvard University,
60 Garden Street, $MS-51$, Cambridge, MA, 02138 U.S.A.}
\email{anastasia.fialkov@cfa.harvard.edu}
\email{aloeb@cfa.harvard.edu}

\begin{abstract}
The dispersion measure of extragalactic radio  transients, such as of recently discovered Fast Radio Burst  FRB150418, can be used to measure the column density of free electrons in the  intergalactic medium. The same electrons  also scatter the Cosmic Microwave Background (CMB) photons, affecting precision measurements of  cosmological parameters. We explore the connection between the dispersion measure of  radio transients existing during the Epoch of Reionization (EoR) and the total optical depth for the CMB, $\tau_{CMB}$, showing that the existence of such transients would  provide a new sensitive probe of $\tau_{CMB}$. As an example, we consider the population of FRBs. Assuming they exist during the EoR, we show that:  (i) such sources can probe the reionization history by measuring $\tau_{CMB}$ to sub-percent accuracy, and  (ii) they can be detected with high significance by an instrument such as the Square Kilometer Array.
\end{abstract}


\section{Introduction}

The Epoch of Reionization (EoR)  is one of the major research areas in cosmology today \citep{Loebbook}. It started around $z\sim13$, and by redshift of $z\sim 6$ most of the  hydrogen atoms in the intergalactic medium were ionized by the UV photons emitted by stars and quasars \citep{Zahn:2012,George:2015,Ade:2015,Becker:2015}. Hydrogen reionization was closely followed by first helium reionization during which helium atoms lost their outer electron, while second helium reionization occurred around $z\sim 3$. The reionization of the intergalactic medium (IGM) is a nuisance for the  Cosmic Microwave Background (CMB) cosmology as CMB  photons scatter off the ionized gas. The precision with which the total optical depth to reionization, $\tau_{CMB}$, can be measured from the  CMB data is very poor, e.g., the 68\% confidence level in $\tau_{CMB}$ corresponds to a relative error of $\sim 24\%$ \citep{Ade:2015}. Because so little is known about the EoR at present, the large uncertainty in  $\tau_{CMB}$ propagates to other cosmological parameters extracted from the CMB data. Luckily, alternative probes of reionization, such as the 21-cm line of hydrogen, will in the future  provide independent constraints on $\tau_{CMB}$ and help to remove the related uncertainty \citep{Liu:2015, Fialkov:2016}.

In addition to the CMB scattering, the ionized plasma leaves other fingerprints on observed radiation. Any radio waves will follow the dispersion relation  $w^2=w_p^2+k^2c^2$  as they propagate  through cold unmagnetized plasma with electron density $n_e$, where  $w = 2\pi\nu$ with $\nu$ being the  frequency of the wave and $k$ its wavenumber, $c$ is the speed of light, and $w_p = 5.641\times 10^4~(n_e/{\rm cm}^{-3})^{1/2}$ s$^{-1}$  is the plasma frequency. As a result, the group velocity of any radio wave packet is modified in a frequency-dependent way as the signal travels through the ionized plasma. In such a medium, a transient at frequency $\nu$ is delayed by  $\Delta t = 4.15\times10^{-3}\textrm{DM}/\nu^2$ seconds,  where $\nu$ is in GHz and the dispersion measure, DM,  is defined as the integrated electron density along the line of sight in  units of pc cm$^{-3}$.  For extragalactic sources, the  DM includes the contributions from  the interstellar medium of the Milky Way,  the host galaxy of the source, and the  IGM. For sources at high redshifts ($z\gtrsim 0.5$), the ionized IGM should be the dominant source of dispersion (DM $\gtrsim 500$ pc cm$^{-3}$) without substantial contamination from the Milky Way,  the local Universe and the host galaxy (DM $\lesssim 250$ pc cm$^{-3}$). Thus, measuring the dispersion of high-redshift sources offers a unique way to probe the ionized IGM and constrain the cosmic reionization history.  In particular, the dispersion of radio emission by high redshift Gamma Ray Bursts (GRB) was discussed  as a probe of the IGM by \citet{Ginzburg:1973}, \citet{Palmer:1993}, \citet{Ioka:2003} and \citet{Inoue:2004}, and the recently discovered Fast Radio Burst (FRB) FRB150418, associated with a galaxy at $z\sim 0.492$, provided a direct measurement of the cosmic baryon density  of the IGM of $4.9\%$ through its DM measurement \citep{Keane:2016}.

Although both the dispersion measure and  $\tau_{CMB}$ probe the same column density of the free electrons, the connection between these two quantities has not been made in the literature. Assuming that radio transients exist at high enough redshifts, we  discuss their cosmological implications in Section \ref{Sec:TDM} focusing on the relation between $\tau_{CMB}$ and DM and showing that an ensemble of high-redshift radio transients would probe  $\tau_{CMB}$ to sub-percent accuracy.  In Section \ref{Sec:SKA}  we discuss perspectives for such measurement. Assuming that  radio transients, such as the population of FRBs,  exist at high enough redshifts, we show that  the Square Kilometer Array\footnote{https://www.skatelescope.org/} (SKA) will have sufficient sensitivity to probe these sources out to $z\sim 14$ and fully constrain the history of reionization through their DM.  We conclude in Section \ref{Sec:conc}. Throughout this work we use cosmological parameters  $H_0$, $Y_P$, $\Omega_b$, $\Omega_m$ and $\Omega_\Lambda$ from \citet{Ade:2015}.

\section{$\tau$-DM relation} 
\label{Sec:TDM}

We start by exploring the connection between the  dispersion measure in the signal of a cosmological transient at a redshift $z$,
\begin{equation}
{\rm DM}(z) = \int_0^{z}  \frac{n_e(z')}{1+z'} dl,
\label{Eq:DM}
\end{equation}  
    and  the contribution of the IGM at redshifts out to $z$ to the total optical depth of the  CMB,
\begin{equation}
\tau(z)  = \int_0^{z} \sigma_T n_e(z') dl,
\label{Eq:Tau}
\end{equation}
where $\sigma_T = 6.25\times10^{-25}$ cm$^2$ is the Thompson cross-section, $dl=cdt = -cdz'/(1+z')H(z')$  is the differential of the proper distance and $H(z) = H_0\left[\Omega_m(1+z)^3+\Omega_\Lambda\right]^{1/2}$ is the Hubble parameter. Taking $z$ to the beginning of the EoR provides the total optical depth of the CMB, $\tau_{CMB}$. Both DM and $\tau(z)$ depend on the redshift evolution of electron density
\begin{equation}
n_e(z) = \bar n_{b}(z)\left[(1-Y_P)x_{HII} +\frac{Y_P}{4}\left(x_{HeII} +x_{HeIII}\right)\right],
\end{equation}
where $\bar n_{b}(z) = \Omega_b\rho_{cr,0}(1+z)^3/m_p$ is the mean baryon density at redshift $z$,  $\rho_{cr,0}$ is the critical density today,   $m_p$ is the proton mass, and $Y_P$ is the helium mass fraction. The redshift evolution of the electron density depends on the details of reionization. In the above equation  $x_{HII}$ is the redshift-dependent hydrogen ionization fraction, $x_{HeII}$ is the fraction of singly ionized helium and $x_{HeIII}$ is the fraction of fully ionized helium. 

\begin{figure*}
\includegraphics[width=2.35in]{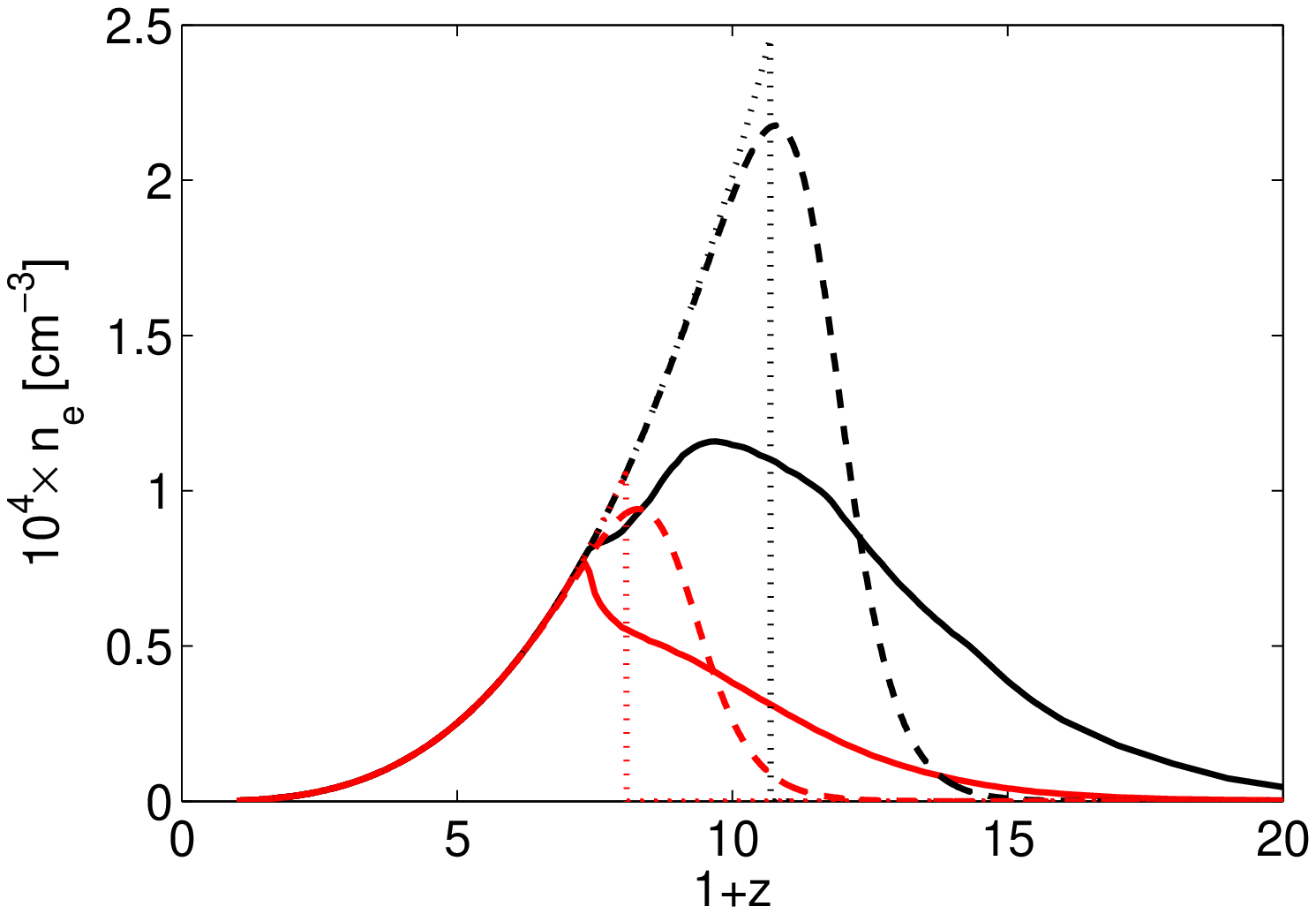}\includegraphics[width=2.35in]{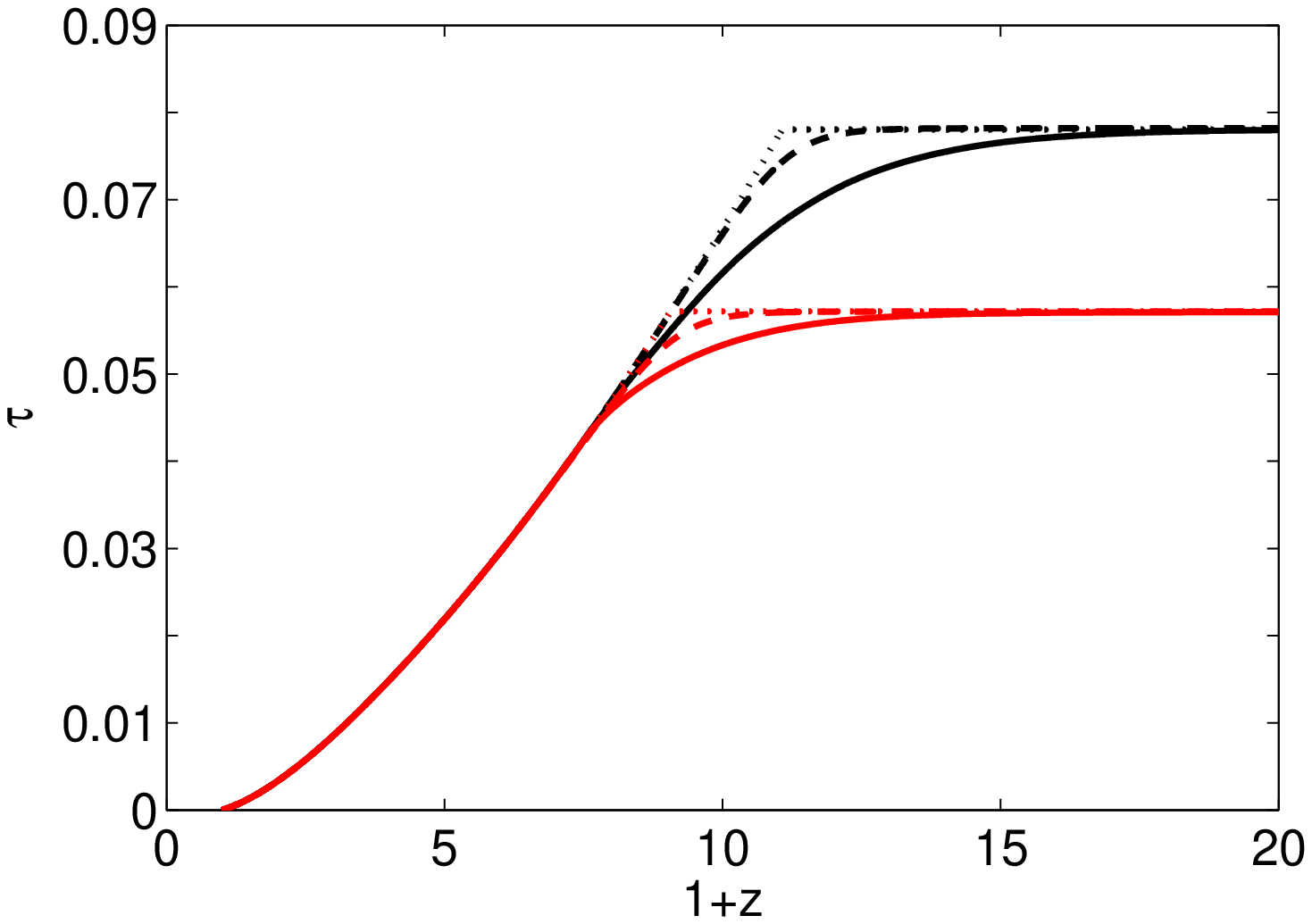}\includegraphics[width=2.35in]{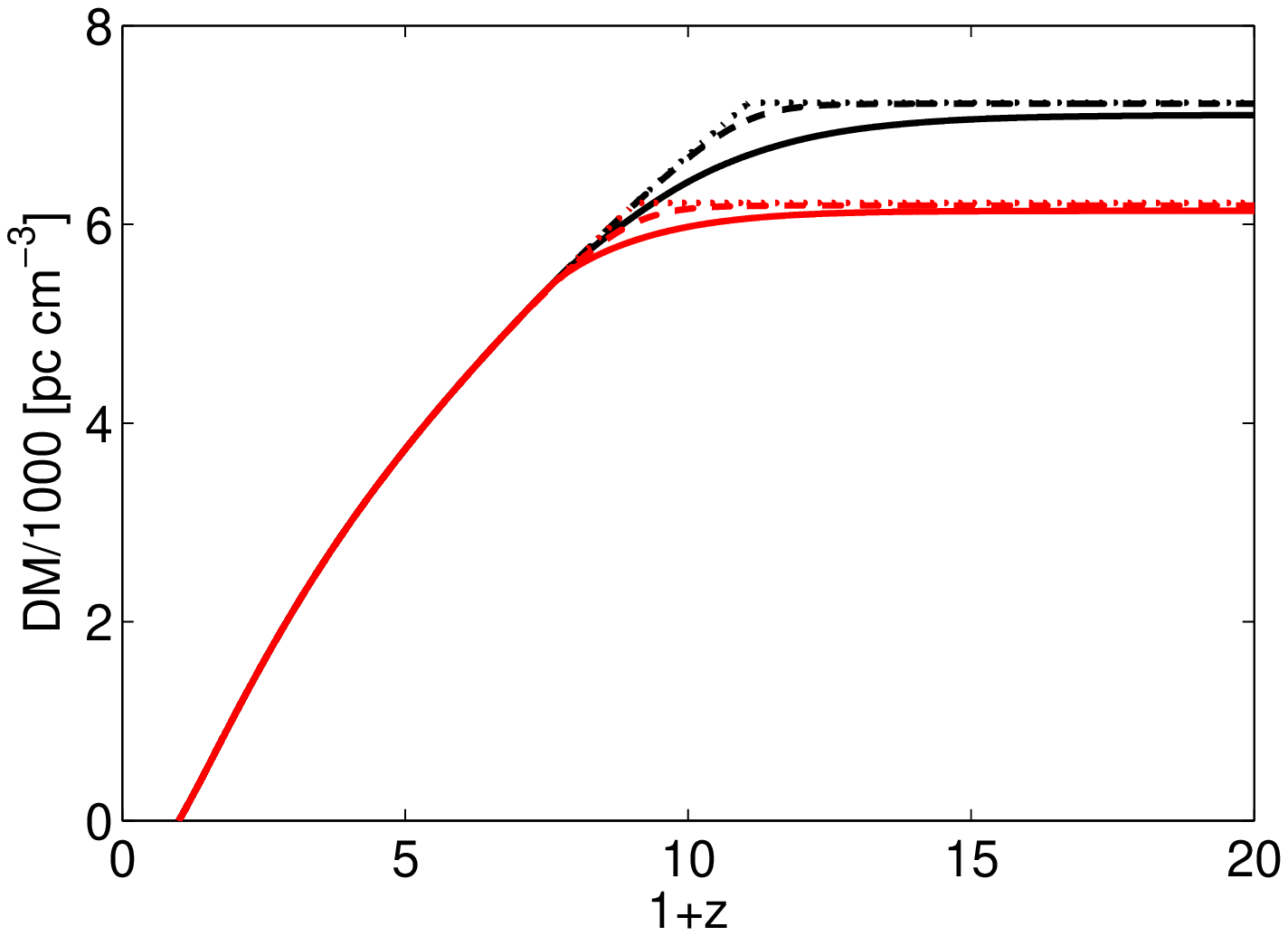}
\caption{Rescaled free electron density, $10^4\times n_e(z)$ [cm$^{-3}$], at redshift $z$ (left), integrated CMB optical depth out to redshift $z$ (middle panel) and rescaled dispersion measure (DM/1000 [pc cm$^{-3}$]) for a source at redshift $z$ (right)  are shown for a realistic reionization scenarion (solid), $\tanh$ model of reionization (dashed) and an instanteneous reionization (dotted) for $\tau = 0.057$ (red) and $\tau = 0.078$ (black).   }
\label{Fig:1}
\end{figure*}

In the left panel of Figure \ref{Fig:1} we show examples of $n_e(z)$ for three different models of the reionization history each shown for two values of $\tau_{CMB}$ = 0.057 and 0.078, both within $1-\sigma$ confidence level of the recent  Planck satellite measurement $\tau_{CMB} = 0.066\pm0.016$ \citep{Ade:2015}.  Assuming that hydrogen and first helium reionization occur simultaneously due to their similar ionization potentials (i.e., $x_{HII} = x_{HeII}$), we adopt the following models (i) an instantaneous reionization where the hydrogen ionization fraction jumps from 0 to 1 at $z_{re}= 8.1$ (10.7) for $\tau_{CMB}$ = 0.057 (0.078), (ii) a reionization history  parameterized by the conventional form 
\begin{equation}
x_{HII} = \frac{1}{2}\left[1-\tanh\left(\frac{z-z_{re}}{\Delta z_{re}}\right)\right]
\label{Eq:xHII}
\end{equation}
for the same values of $z_{re}$ and $\Delta z_{re} =  1$, and (iii) a realistic reionization history which for $\tau_{CMB}$ = 0.078 assumes that star formation occurs in halos with mass larger than the atomic cooling threshold, M$_{\rm h}\gtrsim 10^8\left[(1+z)/10\right]^{-3/2}$ M$_\odot$,  and accounts for the effect of the photoheating feedback on star formation \citep{Cohen:2015}, while in the case of  $\tau_{CMB}$ = 0.057 only very massive halos contributed to reionization, M$_{\rm h}\gtrsim 10^9\left[(1+z)/10\right]^{-3/2}$ M$_\odot$. Finally, we assume a simple model for second helium reionization with $HeIII$   fraction  being of the same form as in Eq. (\ref{Eq:xHII}) with $z_{re} = 3$ and $\Delta z_{re} = 1$.

Figure \ref{Fig:1} also shows the integrated optical depth out to redshift $z$ (middle panel) and the DM of a source at redshift $z$ (right panel) for the three reionization  histories and the two values of $\tau_{CMB}$.  Both DM and $\tau$ saturate at high redshifts before the beginning of reionization when the free electron fraction drops to zero. Depending on the normalization, the value to which $\tau(z)$ saturates is either $\tau_{CMB} = 0.078$ or $\tau_{CMB} = 0.057$, while the upper limit on the DM is slightly different for each reionization history. For the high optical depth the DM takes the values of 7225 pc cm$^{-3}$ in the case of instanteneous EoR, 7213  pc cm$^{-3}$ for the  $\tanh$ model  and 7101 pc cm$^{-3}$ for the more realistic scenario; while for the low optical depth the values are 6215, 6190 and 6137 pc cm$^{-3}$ respectively.

\begin{figure}
\includegraphics[width=3.4in]{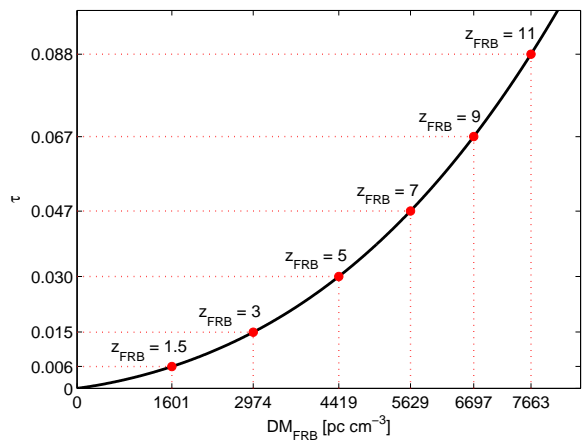}
\caption{$\tau$ as a function of the DM in a fully ionized IGM. For a transient at redshift $z_{FRB}$ the horizontal and vertical dotted lines mark  $\tau(z_{FRB})$ and the DM of the transient respectively.}
\label{Fig:2}
\end{figure}

Although the dependence of the dispersion measure of high-redshift transients on the reionization history has been explored in literature \citep{Ioka:2003, Inoue:2004}, the connection between the intergalactic DM   and  $\tau(z)$ has not been made. Given a series of the DM measurements out to  redshift $z$ for an extragalactic transient population, $\tau(z)$ can be computed precisely.   Using integration by parts we can analytically derive the relation between $\tau(z)$ and DM$(z)$
\begin{equation}
\tau(z) = \left[\frac{{\rm DM}(z)}{{\rm cm}^{-2}}(1+z)-\int_0^{z}\frac{{\rm DM}(z')}{{\rm cm}^{-2}}dz'\right]\times \sigma_T.
\label{Eq:TauDM}
\end{equation}
An example for $\tau$ as a function of the DM is shown in Figure \ref{Fig:2}. Probing the evolution of the DM with redshifts gives a measure of $\tau(z)$ and integrating over the whole range of the DM up to its maximal value should allow one to derive $\tau_{CMB}$. The shape of the $\tau({\rm DM})$ curve has a weak dependence on the  reionization history. For the same  $\tanh$ model  of EoR, changing the duration of reionization from $\Delta z_{re} =1$ to $\Delta z_{re} = 4$ decreases the value of the DM by $\sim 1.4\%$  which is of order the contribution from the interstellar medium inside the Milky Way to the total DM and is small compared to the extragalactic contribution for cosmological transients.

The error in the determination of the optical depth can be expressed as 
\begin{equation}
\Delta \tau(z) = \left[\frac{\Delta {\rm DM}(z)}{{\rm cm}^{-2}}(1+z)-\int_0^{z}\frac{\Delta {\rm DM}(z')}{{\rm cm}^{-2}}dz'\right]\times \sigma_T
\end{equation}
where to leading order the uncertainty due to the redshift error $\Delta z$ cancels out. If  cosmological transients are detected during the EoR this method is expected to provide $\tau_{CMB}$  measurements to a good precision. Assuming a constant error of  $\Delta $DM = 100  pc cm$^{-3}$  (typical uncertainty due to host galaxies) we get $\Delta \tau/\tau \sim  0.3\%$.

\section{Observational prospects}
\label{Sec:SKA}

There is no doubt that probing the reionization history with highly dispersed transients is very challenging, requiring:  (i) bright  radio transients at high enough redshifts, and  (ii) sensitive enough telescope which is designed to probe the corresponding time delays. Here we consider a possibility that recently discovered FRBs could play the role of cosmological beacons and regard the future SKA as a tool to probe them.

\subsection{FRBs as Cosmological Beacons}
FRBs are short ($\sim$ millisecond), bright ($\sim$ Jy) pulses at radio frequencies ($\sim$ 1 GHz) observed over the entire sky \citep{Katz:2016, Petroff:2016, Keane:2016}.  Out of 17 known FRBs \citep{Petroff:2016, Keane:2016} 15 bursts, including the first FRB discovered in 2007 \citep{Lorimer:2007},   were  found in the data collected by  Parkes Radio Telescope in the radio band with central frequency of  $\sim 1372.5$ MHz and bandwidth of 288 MHz \citep{Lorimer:2007, Keane:2011, Thornton:2013, Burke-Spolaor:2014, Champion:2015, Petroff:2015, Ravi:2015, Keane:2016}, one FRB was detected in the Arecibo Pulsar ALFA Survey at $1225-1525$ MHz frequencies \citep{Spitler:2014} and one FRB was found in the Green Bank Telescope (GBT) data within  the 700-900 MHz frequency range \citep{Masui:2015}. These bursts are   very bright pulses with the peak detected flux, $S_{\textrm{peak}}$, being in the 0.3-1.3 Jy range for all the bursts  except for the original ``Lorimer'' burst    which had $S_{\textrm{peak}}>30$ Jy  \citep{Lorimer:2007}. Except for recently discovered FRB150418  which was followed by a $\sim$6 day radio afterglow, the radio bursts observed so far are not known to be accompanied by emission in other wavelength regimes, such as X-rays or gamma-rays, making their identification more difficult.

The short duration of FRBs   allows one to probe the delay time  at each frequency. As for pulsars, the observed delay  scales as $\nu^{-2}$ \citep{Keane:2011, Thornton:2013, Burke-Spolaor:2014, Masui:2015, Petroff:2015}, which can be explained by  dispersion in a cold plasma. However, the typical DM for FRBs, ${\rm DM}= 375-1629$ pc cm$^{-3}$, is much larger than what is usually measured for the Galactic transients, ${\rm DM}\lesssim 250$ pc cm$^{-3}$. Therefore, an extragalactic explanation for the observed FRBs,  which was recently supported by the host galaxy identification for  FRB150418 \citep{Keane:2016},  is usually invoked \citep{Lyubarsky:2014, Totani:2013, Zhang:2014, Geng:2015, Falcke:2014} with the observed DM of FRBs corresponding to their redshifts in the range $z\sim 0.2-1.4$. However, there might be several distinct populations of sources contributing to the discovered FRBs, and there are theoretical models where  the observed high DM of FRBs is be produced locally \citep{Kulkarni:2014, Loeb:2014}.

In the following we consider a population of FRBs similar to the detected ones but located at much higher  redshifts during the EoR and consider prospects for their detection. Using the existing data  we first estimate the redshift of each known FRB, $z_0$, relying on the observed dispersion measure  \citep{Petroff:2015}. For our purposes, the DM contributions from both the Milky Way and the host FRB galaxy are  considered to be foregrounds, which we  remove to isolate the cosmological signal. We estimate the extragalactic contribution  as  DM$_{\textrm {EG}}$ = DM$_{\textrm {Measured}}$- DM$_{\textrm {MW}}$, ignoring the host contribution which scales as $(1+z)^{-1}$ for cosmological FRBs. The Milky Way contribution based on the  NE2001 Galactic electron density model \citep{Cordes:2002} is given by \citet{Petroff:2015} for each FRB location. 

Next, we estimate the  intrinsic luminosity of the bursts at their rest-frame central frequencies $\nu_0 = \bar \nu (1+z)$
\begin{equation}
L_{\nu_0}^{\textrm {peak}} = \frac{4\pi D_L^2(z_0) S_{\bar \nu}^{\textrm {peak}}}{(1+z_0)},
\label{Eq:L}
\end{equation}
where $\bar \nu$ is the mean observed frequency,  $D_L(z_0)$ is the luminosity distance,  assuming the source emission is isotropic. We also assume a flat spectral energy distribution since there is no spectral information available for  most of the detected FRBs with the exception of the burst detected in the Arecibo  Survey  \citep{Spitler:2014} and the recently detected FRB150418. Finally, we choose four representative FRBs out of the observed ones selecting an FRB with the Maximal (excluding the ``Lorimer'' burst), Median and Minimal intrinsic luminosity densities and also considering the the ``Lorimer'' burst. The details ($z_0, \nu_0, L_{\nu_0}^{\textrm {peak}}$) of these bursts are summarized in Table 1. 

\begin{deluxetable*}{lllllll}
\tabletypesize{\scriptsize} \tablecaption{Selected FRBs from the catalog of \citet{Petroff:2016} with the  minimal, median, and maximal (excluding the Lorimer burst) intrinsic peak differential luminosity and the Lorimer burst. } \tablewidth{0pt}
\tablehead{  \colhead{Type}  &  \colhead{Event} & \colhead{Telescope}  & \colhead{DM$_{\textrm {EG}}$ [pc cm$^{-3}$]} & \colhead{$z_0$} & \colhead{$\nu_0$ [GHz]} & \colhead{$L_{\nu_0}^{\textrm {peak}}$ [erg s$^{-1}$ Hz$^{-1}$ sr$^{-1}$]}}
\startdata 
Min & FRB010621 \citep{Keane:2011} & Parkes & 748 & 0.22 & 1.7  & $5.2\times 10^{32}$\\ 
Median & FRB110523 \citep{Masui:2015}  & GBT & 623 & 0.55 & 1.2  & $5.0\times 10^{33}$\\ 
Max & FRB110220 \citep{Thornton:2013} & Parkes & 944 & 0.85 & 2.55  & $2.6\times 10^{34}$\\ 
Lorimer & FRB010724 \cite{Lorimer:2007}  & Parkes & 375 & 0.32 & 1.86  & $>8.2\times 10^{34}$\\ 
\enddata
\label{Tab:1}
\end{deluxetable*}

\subsection{Prospects of Observation with SKA}
\begin{figure*}
\includegraphics[width=3.4in]{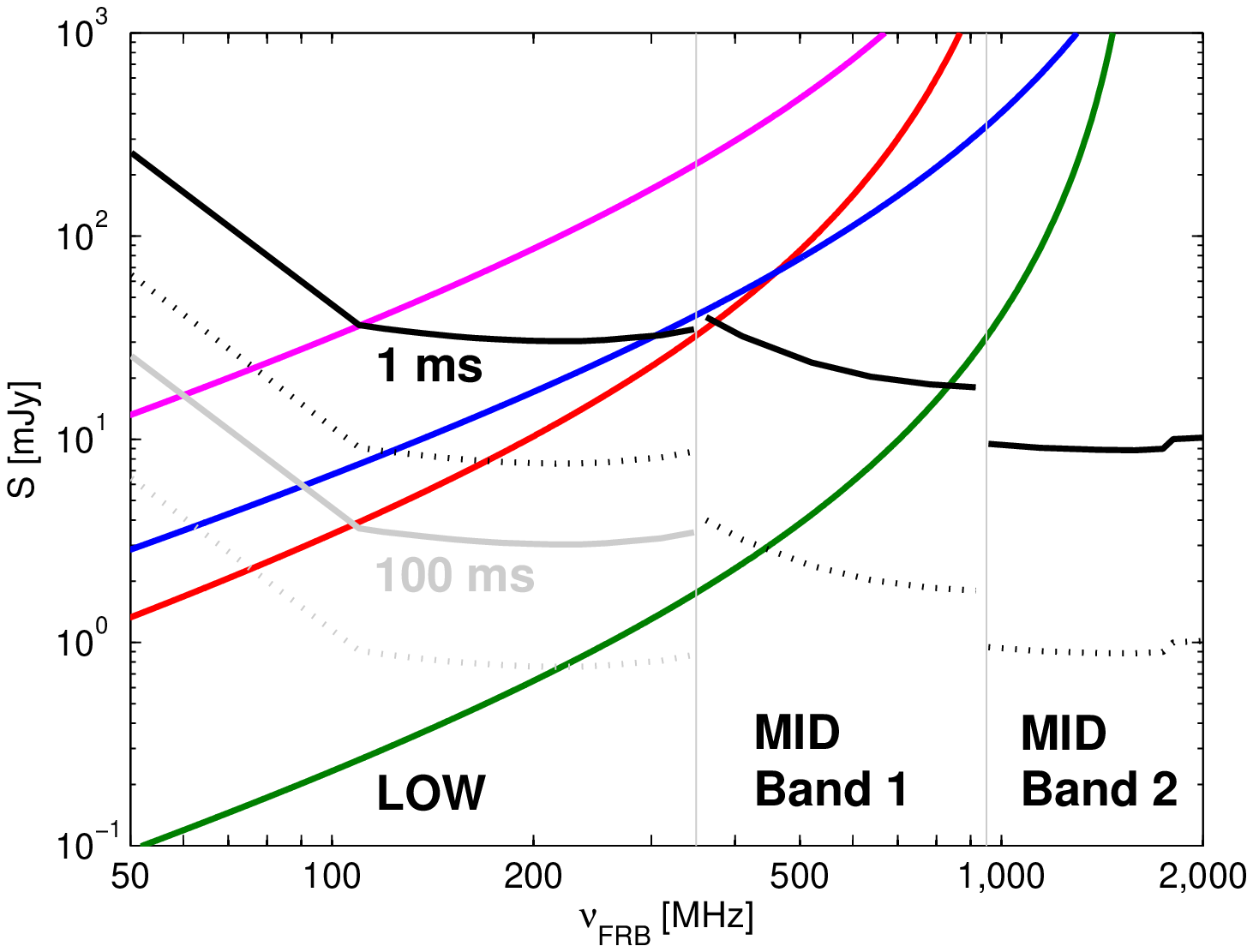}\includegraphics[width=3.4in]{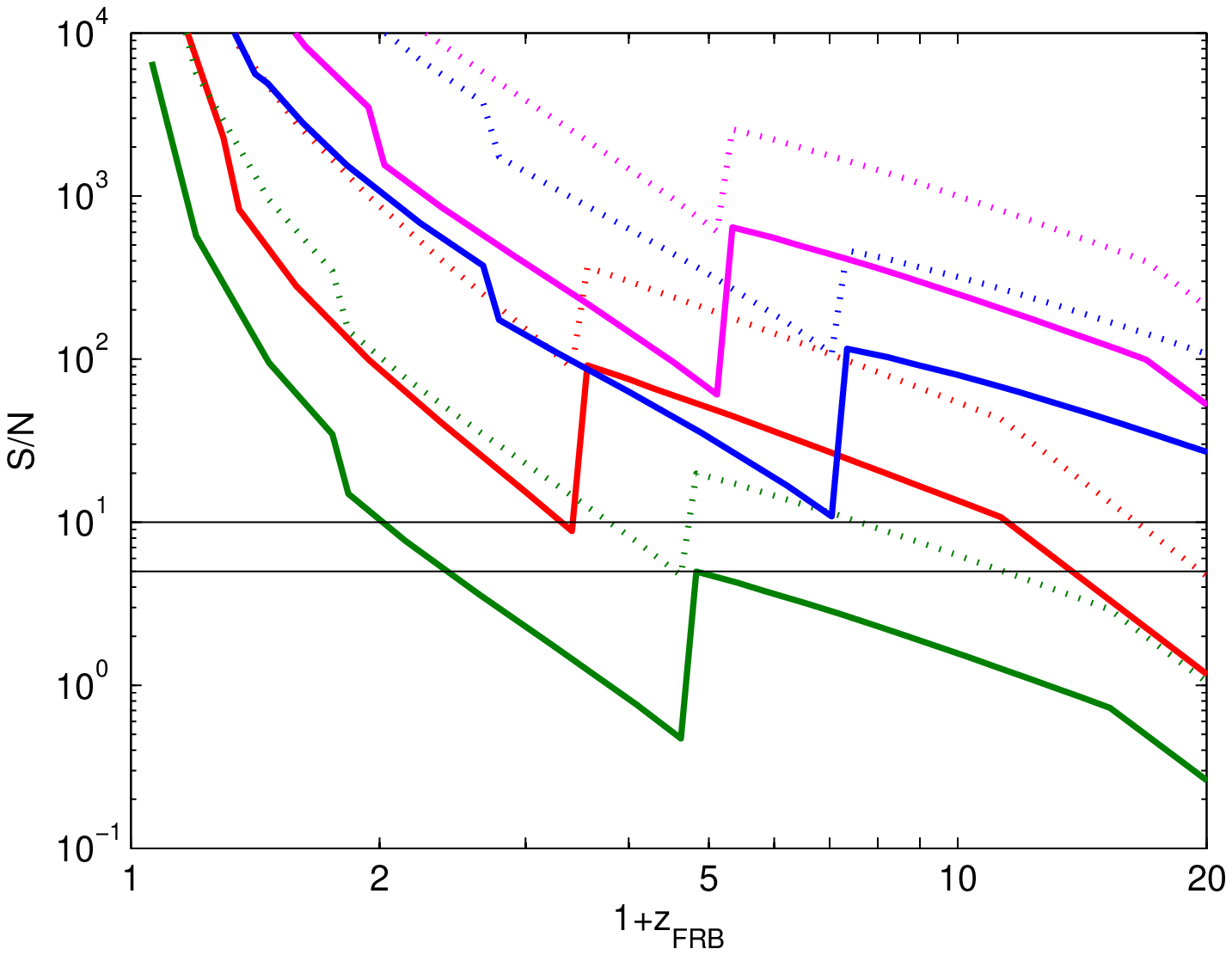}
\caption{Left: Prospect of detection for the four selected FRBs listed in Table 1 placed at different cosmological distances. We show  Minimal (green),  Median (red), Maximal (blue) and Lorimer (magenta), together with the  5-$\sigma$ noise level of Phase 1 (solid) and Phase 2 (dotted) of SKA-MID with 1 ms integration time,  and SKA-LOW  with 1 ms  (black) and 100 ms (grey) integration times.  Right: Signal to noise ratio for each FRB as a function of its redshift. We show the results for Phase 1 (solid) and Phase 2 (dotted) of SKA for  Minimal (green), Median (red), Maximal (blue) and Lorimer (magenta) FRBs. We set an integration time of  1 ms for SKA-MID and 100 ms for SKA-LOW. Horizontal lines mark 10 and 5  signal-to-noise thresholds.  }
\label{Fig:SN}
\end{figure*}

We consider the four representative FRBs (Table 1) as four types of sources with different characteristic luminosities, i.e., with their intrinsic luminosity $L_{\nu_0}^{\textrm {peak}}$ fixed,  and estimate the signal to noise with which each type can be probed with the SKA. For each burst we reverse the Eq. (\ref{Eq:L}) and find the peak flux of each  FRB varying its redshift $z_{FRB}$, or, equivalently, its observed frequency  $\nu_{FRB}$.  The expected scaling of the  signal with  $\nu_{FRB}$ is shown on the left panel of Figure \ref{Fig:SN} for each type of the FRBs together with the expected $10\sigma$ sensitivity of SKA at each frequency shown for Phase 1 and 2 of the SKA-MID in Band 1 ($0.35-0.95$  GHz) and Band 2 ($0.95-1.76$ GHz) and of SKA-LOW ($50-350$ MHz). For both SKA-MID, which is an array of dish antennas, and SKA-LOW,  an  array of low-frequency dipole  antennas, the  sensitivity is estimated as follows
\begin{equation}
\sigma = \frac{\textrm{SEFD}}{\sqrt{2 t_{\textrm{int}} \Delta \nu_{\textrm{BW}}}},
\end{equation}
where $t_{\textrm{int}}$ is the integration time, $\Delta \nu_{\textrm{BW}}$ is the bandwidth and factor 2 of accounts for 2 polarization channels. SEFD $\equiv 2 k_B\left(A_e/T_{sys}\right)^{-1}$ measured in Jy, is the system equivalent flux density which depends on the effective area, $A_e$, and the system temperature, $T_{sys}$,   which for Phase 1 of the SKA can be found in the SKA online documentation\footnote{http://www.skatelescope.org/} for the entire SKA-LOW array and for a single dish of SKA-MID. To estimate the sensitivity of Phase 1 of SKA-MID we assume that the effective area of the array is 133 times the area of a single dish. Based on the SKA online documentation we assume that Phase 2 of SKA-MID  will be 10 times better than Phase 1 in terms of SEFD,  and SKA-LOW will be 4 times better. On the right panel of Figure \ref{Fig:SN} we show the signal to noise of each FRB as a function of redshift with horizontal lines marking the  $10\sigma$, $5\sigma$ and  $1\sigma$ detection thresholds.

According to our estimates and in agreement with previous studies \citep{Macquart:2015}, SKA-MID will be a very powerful observatory for detecting all the FRBs except for the weakest ones, to which Band-2 of SKA-MID should still be sensitive enough. On the other hand, Phase 1 of SKA-LOW is expected to be sensitive only to the strongest FRBs of the Lorimer type, while  Phase 2 can also be sensitive to the entire brightest half of the FRB population. Increasing integration time can be beneficial in this case since the bursts are stretched in time by the factor of $(1+z_{FRB})$ and the delay time grows for higher redshift sources observed at low frequencies. With different integration times for SKA-MID ($ t_{\textrm{int}}=1$ ms) and SKA-LOW ($t_{\textrm{int}}=100$ ms), which explains the jump in signal to noise,  all FRBs brighter or identical to the Median are detectable with high signal to noise, S/N$>10$, out to $z\sim 11$ and out to $z\sim 14$ with high signal to noise S/N$>5$ even with SKA Phase 1; while weaker FRBs will be detectable with Phase 2 out $z\sim 10.4$  with  S/N$>5$.

With current constraints of the reionization history, FRBs at redshift $\sim 10$ are expected have a very large DM of several thousands (Figure \ref{Fig:1}). To allow for detection of so high DM with a telescope, one simply needs to assume that  the maximal dispersion smearing within a frequency channel, $\Delta t_{\textrm{max}} = 8.3\times 10^{-3} {\rm DM}_{\textrm{max}} \Delta \nu/ \nu_{\textrm{min}}^3 $ seconds, is bigger than the sampling time, $t_{\textrm{samp}}$. Here $\Delta \nu$ is the frequency channel width in GHz, and $\nu_{min}$ is the minimum (lowest) frequency in the observation band. Thus, the maximal DM that can be probed by a telescope is 
\begin{displaymath}
{\rm DM}_{\textrm{max}}   = \frac{t_{\textrm{samp}}  }{8.3\times 10^{-3} \textrm{s}}\frac{\nu_{\textrm{min}}^3}{\Delta \nu}. 
\end{displaymath}
This argument should equally apply for both dishes (such as SKA-MID) and a low-frequency array  (such as SKA-LOW).  Most of the known FRB were detected with cutoff implied on the DM at DM$_{max}\sim$ 2000 pc  cm$^{-3}$ \citep{Masui:2015, Thornton:2013}. However, higher  DMs are being accessed with ongoing surveys.  For example, \citet{Burke-Spolaor:2014} searched DM trials up to 3000 pc  cm$^{-3}$; while a recently developed pipeline allows the data  searched over the DM  from 0 to 5000  pc  cm$^{-3}$ and over widths  from 0.128 to 262 ms  \citep{Keane:2015b}. However, no FRBs with the DM above 1700 pc  cm$^{-3}$ were detected so far.  Future instruments, such as the SKA,  will allow one to explore higher values of the DM. In particular, with the Phase 1 of the SKA  the search out to dispersion measures of 3000 pc cm$^{-3}$ is proposed allowing to probe FRBs out to $z\sim 3$,  while with SKA Phase-2 it should be possible to search out to dispersion measures of 10,000  pc cm$^{-3}$ and beyond \citep{Keane:2015}  probing the FRB population during the EoR.

In the above discussion we assumed that there is no  significant scattering in the intergalactic medium. With  scattering, the observed flux is expected to decrease \citep{Hassall:2013}. However, for all the observed FRBs the scattering by the IGM was estimated  to be small \citep{Thornton:2013}. Taking this into  account together with the fact that the peak flux could have been missed by observation, the signal which we quoted above (and, thus, the signal-to-noise values) should be regarded as lower limits only.

\section{Conclusions}
\label{Sec:conc}

In this Letter we explored the connection between the DM of high-redshift radio transients  and the total CMB optical depth $\tau_{CMB}$ by deriving an analytic $\tau$-DM relation, which can be used to constrain $\tau_{CMB}$ provided radio transients exist during reionization.  Assuming sources, such as FRBs,  exist at high redshifts, we have shown that  Phase 1 of the SKA or a similar radio telescope should be able to probe the brightest half of the population out to high redshifts, $z\sim 14$. In this case, it should be possible to test the full reionization history with Phase 1 of the SKA through DM of these sources. According to the latest estimates, FRBs are frequent events, with  $\sim 6000$ of them occurring every day  \citep{Champion:2015}, and at least one of them was confirmed to be extragalactic \citep{Keane:2016}. Therefore, there are good chances for  detection of more FRBs at cosmological distances with future telescopes, such as the SKA, especially when exploring new region of the parameter space, such as higher dispersion measures and lower frequencies. Moreover, depending on the luminosity function of the population, high redshift sources could be gravitationally lensed by the foreground large scale structure, and, thus, more easily observed \citep{Wyithe:2011, Fialkov:2015}.

\acknowledgments

We thank Mark Reid and Jim Moran for useful discussions. This work was supported in part NSF grant AST-1312034 (for A.L.).


\begin{thebibliography}{}

\bibitem[Ade et al. (2015)]{Ade:2015}  Planck Collaboration; Ade, P. A. R., et al. 2015, arXiv:1502.01589


\bibitem[Becker et al. (2015)]{Becker:2015} Becker, G. D., Bolton, J. S., Madau, P., et al. 2015, MNRAS,
447, 3402

\bibitem[Burke-Spolaor \&  Bannister (2014)]{Burke-Spolaor:2014}  
Burke-Spolaor, S., \& Bannister, K. W., 2014, ApJ, 792, 19

\bibitem[Champion et al. (2015)]{Champion:2015}  	
Champion, D. J., Petroff, E., Kramer, M., Keith, M. J., Bailes, M., et al., 2015, arXiv:1511.07746

\bibitem[Cohen et al. (2015)]{Cohen:2015}  	 Cohen, A., Fialkov, \& A., Barkana, R., 2015, arXiv:1508.04138

\bibitem[Cordes \& Lazio (2002)]{Cordes:2002} Cordes, J. M. \& Lazio, T. J. W. 2002, arXiv:0207156

\bibitem[Dolag et al. (2015)]{Dolag:2015}
Dolag, K., Gaensler, B. M., Beck, A. M., \& Beck, M. C.
2015, MNRAS, 451, 4277


\bibitem[Falcke \& Rezzolla (2014)]{Falcke:2014}
Falcke, H. \& Rezzolla, L. 2014, A\&A, 562, 137

\bibitem[Fialkov \& Loeb (2016)]{Fialkov:2016}
Fialkov, A. \& Loeb, A.,  2016, arXiv:1601.03058

\bibitem[Fialkov \& Loeb (2015)]{Fialkov:2015}
Fialkov, A. \& Loeb, A.,  2015, ApJ, 806, 256

\bibitem[Geng \& Huang (2015)]{Geng:2015}
Geng, J. J. \& Huang, Y. F. 2015, ApJ, 809, 24

 \bibitem[\protect\citeauthoryear{George}{2015}]{George:2015}
George, E. M., Reichardt, C. L., Aird, K. A., Benson, B. A., Bleem, L. E., et al. 2015, ApJ, 799, 177 


\bibitem[Ginzburg (1973)]{Ginzburg:1973}
Ginzburg V.L., 1973, Nature, 246, 415

\bibitem[Hassall et al. (2013)]{Hassall:2013}  		
Hassall, T. E., Keane, E. F., \& Fender, R. P., 2013, MNRAS, 436, 317 

\bibitem[Ioka (2003)]{Ioka:2003} 
 Ioka, K., 2003, ApJ 598, L79
 
 \bibitem[Inoue (2004)]{Inoue:2004} 
  Inoue, S., 2004, MNRAS 348, 999



\bibitem[Keane et al. (2011)]{Keane:2011}
Keane, E. F., Kramer, M., Lyne, A. G., Stappers, B. W., McLaughlin, M. A., 2011, MNRAS, 415, 3065

\bibitem[Keane et al. (2015)]{Keane:2015}
Keane, E., Bhattacharyya, B., Kramer, M., Stappers, B., Keane, E. F., et al., 2015, ''A Cosmic Census of Radio Pulsars with the SKA``, Proceedings of Advancing Astrophysics with the Square Kilometre Array (AASKA14). 9 -13 June, 2014. Giardini Naxos, Italy. 

\bibitem[Keane \& Petroff (2015)]{Keane:2015b}
Keane, E. \& Petroff, E.,  2015, MNRAS, 447, 2858

\bibitem[Keane et al. (2016)]{Keane:2016}
Keane, E. F., Johnston, S., Bhandari, S., Barr, E., Bhat, N. D. R., et al., 2016, Nature, 17140

\bibitem[Kulkarni et al. (2014)]{Kulkarni:2014}
Kulkarni, S. R., Ofek, E. O., Neill, J. D., Zheng, Z., \& Juric, M.
2014, ApJ, 797, 70

\bibitem[Katz (2016)]{Katz:2016}
Katz, J. I., arXiv:1505.06220


\bibitem[Liu et al. (2015)]{Liu:2015}
Liu, A., Pritchard, J. R., Allison, R., Parsons, A. R., Seljak, U., Sherwin, B. D., 2015, arXiv:1509.08463

\bibitem[Loeb \& Furlanetto (2013)]{Loebbook} Loeb, A. \& Furlanetto, S.  2013, The First Galaxies in the Universe, Princeton University Press   (Princeton)

\bibitem[Loeb et al. (2014)]{Loeb:2014}
Loeb, A., Shvartzvald, Y., Maoz, D., 2014, MNRAS, 439, 46	


\bibitem[Lorimer et al. (2007)]{Lorimer:2007}
Lorimer, D. R., Bailes, M., McLaughlin, M. A., Narkevic, D. J., \& Crawford,
F. 2007, Sci, 318, 777



\bibitem[Lyubarsky (2014)]{Lyubarsky:2014}
Lyubarsky, Y. 2014, MNRAS, 442, L9

\bibitem[Macquart et al. (2015)]{Macquart:2015}
Macquart, J. P., Keane, E., Grainge, K., McQuinn, M., Fender, R., et al., 2015, Proceedings of Advancing Astrophysics with the Square Kilometre Array (AASKA14). 9 -13 June, 2014. Giardini Naxos, Italy.

\bibitem[Masui et al. (2015)]{Masui:2015}
Masui, K., Lin, H.-H., Sievers, J., Anderson, C. J., Chang, T.-C., et al., 2015, Nature, 528, 523

\bibitem[Mingarelli et al. (2015)]{Mingarelli:2015}
Mingarelli, C. M. F., Levin, J., Lazio, T. J. W., 2015, ApJ, 814, 20



\bibitem[Palmer (1993)]{Palmer:1993}
Palmer, D. M., 1993, ApJ, 417, L25

\bibitem[Petroff et al. (2015)]{Petroff:2015}
Petroff, E., Bailes, M., Barr, E. D., Barsdell, B. R., Bhat, N. D. R. et al., 2015, MNRAS, 447, 246


\bibitem[Petroff et al. (2016)]{Petroff:2016}
Petroff, E., Barr, E. D., Jameson, A., Keane, E. F., Bailes, M., et al., 2016, arXiv:1601.03547 

\bibitem[Ravi et al. (2015)]{Ravi:2015}
Ravi V., Shannon R. M., Jameson A., 2015, ApJ, 799, L5

\bibitem[Spitler et al. (2014)]{Spitler:2014}
	Spitler L. G.,  Cordes, J. M., Hessels, J. W. T., Lorimer, D. R., McLaughlin, M. A. et al., 2014, ApJ, 790, 101

\bibitem[Totani (2013)]{Totani:2013}
Totani, T. 2013, PASJ, 65, L12


\bibitem[Thornton et al. (2013)]{Thornton:2013}
Thornton, D., Stappers, B., Bailes, M., et al. 2013, Science, 341, 53

\bibitem[Wyithe et al. (2011)]{Wyithe:2011}
Wyithe, J. S. B., Yan, H., Windhorst, R. A., Mao, S., 2011, Nature, 469, 181

\bibitem[Zahn et al. (2012)]{Zahn:2012} 
Zahn, O., Reichardt, C. L., Shaw, L., Lidz, A., Aird, K. A., et al. 2012, ApJ, 756, 65


\bibitem[Zhang (2014)]{Zhang:2014}
Zhang, B. 2014, ApJ, 780, L21

\end{thebibliography}
\end{document}